\documentclass[reprint,superscriptaddress,amssymb,amsmath,aps,showpacs,10pt,floatfix,prl,longbibliography]{revtex4-2}

\def\AVS{$A$V$_3$Sb$_5$}
\def\Fe3Sn2{Fe$_3$Sn$_2$}
\def\cm{cm$^{-1}$}

\usepackage{graphicx}%
\usepackage{color}
\usepackage{epstopdf}
\usepackage{placeins}
\usepackage{amssymb}
\usepackage{amsmath}
\usepackage{amsfonts}
\usepackage[note-name=, use-sort-key = false]{notes2bib}

\usepackage{xcolor}

\usepackage{color}
\usepackage[colorlinks,bookmarks=false,citecolor=darkblue,linkcolor=red,urlcolor=blue]{hyperref} 
\definecolor{darkred}{rgb}{0.7,0.0,0.0}

\definecolor{darkblue}{rgb}{0,0.02,0.45}

\definecolor{darkgreen}{rgb}{0.02,0.45,0.0}

\definecolor{violet}{rgb}{0.8,0.2,0.6}

\begin{document}
\title{Intriguing Low-Temperature Phase in the Antiferromagnetic Kagome Metal FeGe}

\author{M. Wenzel}
\email{maxim.wenzel@pi1.physik.uni-stuttgart.de}
\affiliation{1. Physikalisches Institut, Universit{\"a}t Stuttgart, 70569 
Stuttgart, Germany}

\author{E. Uykur}
\affiliation{Helmholtz-Zentrum Dresden-Rossendorf, Institute of Ion Beam Physics and Materials Research, 01328 Dresden, Germany}

\author{A. A. Tsirlin}
\affiliation{Felix Bloch Institute for Solid-State Physics, Leipzig University, 04103 Leipzig, Germany}

\author{S. Pal}
\affiliation{1. Physikalisches Institut, Universit{\"a}t Stuttgart, 70569 
Stuttgart, Germany}

\author{R. Mathew Roy}
\affiliation{1. Physikalisches Institut, Universit{\"a}t Stuttgart, 70569 
Stuttgart, Germany}

\author{C. Yi}
\affiliation{Max Planck Institute for Chemical Physics of Solids, 01187 Dresden, Germany}

\author{C. Shekhar}
\affiliation{Max Planck Institute for Chemical Physics of Solids, 01187 Dresden, Germany}

\author{C. Felser}
\affiliation{Max Planck Institute for Chemical Physics of Solids, 01187 Dresden, Germany}

\author{A. V. Pronin}
\affiliation{1. Physikalisches Institut, Universit{\"a}t Stuttgart, 70569 
Stuttgart, Germany}

\author{M. Dressel}
\affiliation{1. Physikalisches Institut, Universit{\"a}t Stuttgart, 70569 
Stuttgart, Germany}
\date{\today}

\begin{abstract}
The properties of kagome metals are governed by the interdependence of band topology and electronic correlations resulting in remarkably rich phase diagrams. Here, we study the temperature evolution of the bulk electronic structure of the antiferromagnetic kagome metal FeGe using infrared spectroscopy. We uncover drastic changes in the low-energy interband absorption at the 100~K structural phase transition that has been linked to a charge-density-wave (CDW) instability. We explain this effect by the minuscule Fe displacement in the kagome plane, which results in parallel bands in the vicinity of the Fermi level. In contrast to conventional CDW materials, however, the spectral weight shifts to low energies, ruling out the opening of a CDW gap in FeGe.
\end{abstract}

\pacs{}
\maketitle
The kagome lattice has been a fruitful playground to study novel quantum phenomena for over 70 years, uniting electronic correlations, topologically nontrivial states, and geometric frustration \cite{Syozi1951, Mekata2003}. Being rooted in the network of corner-sharing triangles, destructive quantum interference results in localized states, i.e., a flat band over the entire Brillouin zone \cite{Kim2023a, Li2018}. Together with the van Hove singularities at the $M$ point, this can lead to exotic electronic orders, including flat-band ferromagnetism, Pomeranchuck instabilities, charge and spin bond order, unconventional superconductivity, nematicity, and Wigner crystallization \cite{Kiesel2013, Ferhat2014, Neupert2022, Pollmann2008, Wu2007, Kiesel2012, Yu2012, Lin2021, Nie2022, Chen2018}.

\begin{figure*}
        \centering
       \includegraphics[width=2\columnwidth]{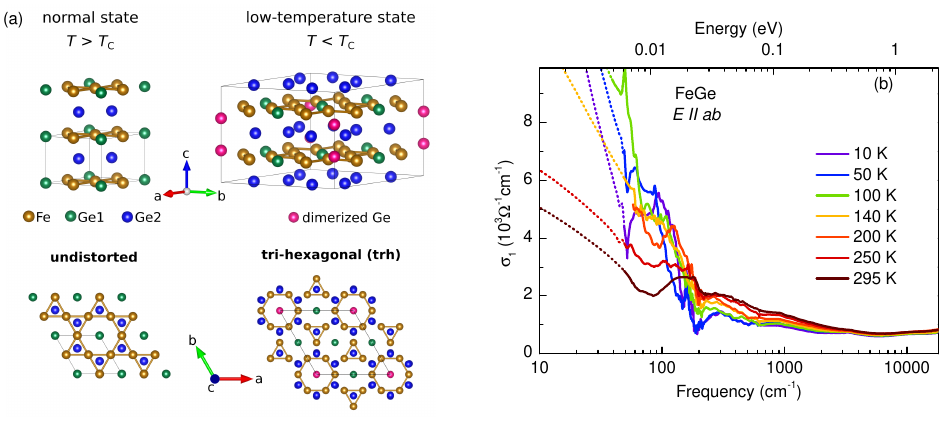}
   \caption{(a) Crystal structure of hexagonal FeGe in the normal- and the low-temperature state. The bottom panel shows the undistorted in-plane Fe-kagome layer, as well as the tri-hexagonal distortion discussed in the literature \cite{Teng2022}. \texttt{VESTA} program was used for the crystal structure visualization \cite{Momma2008}. (b) In-plane optical conductivity calculated from the measured reflectivity at selected temperatures with the dotted lines indicating the Hagen-Rubens extrapolations to low energies. See Fig.~S1~(d) for the spectra at other temperatures. }	
    \label{Fig1}
\end{figure*}

Several of these intriguing phases have been identified in kagome metals, most prominently the exotic charge-density-wave (CDW) in the superconducting \AVS\ ($A$~=~K, Rb, Cs) family \cite{Ortiz2019, Tan2021, Denner2021, Neupert2022, Kang2022, Lin2021}. Recently, charge order in the context of a magnetic kagome metal has been suggested as well \cite{Teng2022, Teng2023, Yin2022}. Hexagonal FeGe crystallizes in the $P6/mmm$ space group featuring magnetic Fe-kagome planes stabilized by Ge1 atoms. Within one unit cell, these kagome layers are separated by a non-magnetic honeycomb lattice of Ge2 atoms as visualized in Fig.~\ref{Fig1}~(a). At room temperature, an out-of-plane antiferromagnetic (AFM) arrangement of the Fe-kagome layers is found, with ferromagnetically ordered Fe-moments within the kagome planes. This AFM order persists down to approximately 60~K, below which a canted AFM arrangement with an additional reorientation at $\sim$~30~K is observed \cite{Bernhard1984, Bernhard1988}.

More recent studies reveal an anomaly at $T_{\mathrm{C}} \simeq 100$~K, where a slight drop in the in-plane magnetic susceptibility occurs \cite{Teng2022, Chen2023, Shi2023}. However, no clear counterpart to this drop is observed in the electric resistivity \cite{Chen2023, Shi2023}. Based on neutron diffraction and scanning tunneling microscopy (STM) measurements, a short-range-ordered CDW phase with a minor in-plane lattice distortion on the order of $10^{-4}$ of the kagome structure has been proposed \cite{Teng2022, Yin2022, Chen2023}. Raman spectroscopy and neutron Larmor diffraction further suggested that deviations from the hexagonal symmetry occur already above $T_{\mathrm{C}}$ \cite{Wu2023a}. 

Although the low-temperature, putative CDW phase of FeGe was investigated extensively by surface-sensitive techniques such as STM and angle-resolved photoemission spectroscopy (ARPES), a systematic study of the temperature-dependent bulk electronic structure remains lacking. Fourier-transform infrared spectroscopy is one of the most effective approaches for investigating bulk CDW order \cite{Huang2013, Ban2017, Vescoli1998, Chen2017, Chen2014, Pfuner2010, Perucchi2004}. Here, the gap opening at the Fermi level below $T_{\mathrm{CDW}}$ is normally manifested in a spectral weight transfer from low to high energies. However, our results on FeGe reveal no indication of a gap opening below $T_{\mathrm{C}}$ and, thus, contradict the CDW scenario at low temperatures. Aided by density functional theory (DFT) calculations of the band structure and the optical conductivity, we show that the reconstruction of the bands due to the structural transition at $T_{\mathrm{C}}$ leads to new interband optical transitions at low energies. We thus interpret the change in the electronic structure below $T_{\mathrm{C}}$ as band splitting rather than gap opening.

High-quality single crystals were synthesized as explained in the Supplemental Material \cite{SM}. Prior to the optical experiments, we measured the in-plane magnetic susceptibility and dc resistivity of our samples. The results are plotted in Fig.~S1~(a), confirming the stoichiometry and identifying the low-temperature anomaly at $T_{\mathrm{C}} = 102$~K. Fig.~\ref{Fig1}~(b) displays the temperature-dependent real part of the in-plane optical conductivity. At first glance, the spectra resemble the optical spectra of other magnetic kagome metals such as the ferromagnetic Fe$_3$Sn$_2$ \cite{Biswas2020, Schilberth2022}, and the ferrimagnetic $R$Mn$_6$Sn$_6$ compounds \cite{Wenzel2022}. However, it is apparent that the low-energy absorptions in FeGe strongly vary with temperature, including a new absorption feature around 100~\cm\ below $T_{\mathrm{C}}$.

Using the Drude-Lorentz approach, different contributions to the total optical spectra are modeled, as shown in Fig.~S2 for selected temperatures. In addition to the classical Lorentzian (interband) and Drude (intraband) contributions, a sharp phonon mode around 190~\cm\ with a Fano-like shape, and a temperature-dependent absorption peak at low energies are observed. The latter feature is assigned to the intraband signature of localized electrons (localization peak), in line with previous optical studies of kagome metals \cite{Wenzel2022a, Uykur2022, Uykur2021, Wenzel2023, Biswas2020, Wenzel2022}. The total complex optical conductivity [$\tilde{\sigma}=\sigma_1 + i\sigma_2$] then takes the form
\begin{equation}
\tilde{\sigma}(\omega)= \tilde{\sigma}_{\mathrm{intraband}} + \tilde{\sigma}_{\mathrm{phonon}} + \tilde{\sigma}_{\mathrm{interband}}.
\label{Cond}
\end{equation}

\begin{figure*}
\includegraphics[width=2\columnwidth]{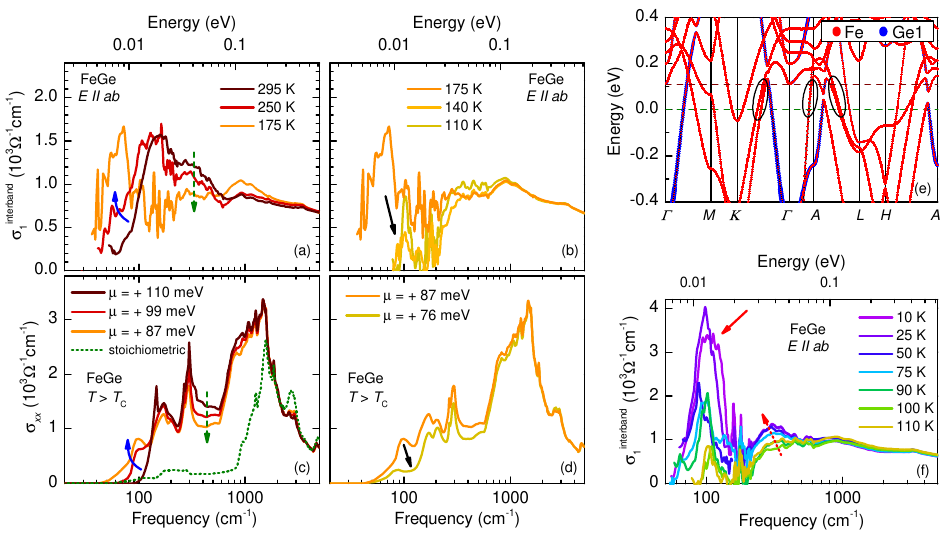}
\caption{(a) and (b) Experimental interband transitions of FeGe at temperatures above $T_{\mathrm{C}}$, obtained by subtracting the Drude and localization peaks and the Fano resonance from the optical conductivity. (c) and (d) DFT in-plane optical conductivities for different electron dopings expressed by the chemical potential $\mu$. The arrows mark characteristic changes in the spectra. (e) Calculated band structure for the high-temperature phase with colored dots symbolizing the contributions of different atoms. The black ovals highlight the linear Fe bands crossing the Fermi level responsible for the low-energy interband optical transitions. Dashed horizontal lines mark the position of the chemical potential for the stoichiometric case (green) and for $\mu = + 110$~meV (brown), corresponding to room temperature. (f) Experimental interband transitions at temperatures below $T_{\mathrm{C}}$. Red arrows mark the enhancement in the spectral weight at low energies due to the structural distortion.}
    \label{Fig2}
\end{figure*}

To further visualize the temperature-driven changes in the low-energy interband absorptions, we subtract the Fano resonance and the intraband contributions, i.e., Drude and localization peaks, from the optical conductivities as presented in Fig.~\ref{Fig2}~(a) and (b) for $T > T_{\mathrm{C}}$. The room temperature data is reproduced by our DFT calculations given in panels (c) and (d) via raising the chemical potential by 110~meV. Upon cooling, the characteristic changes in the experimental spectra can be described by a systematic reduction of the chemical potential. This finding is consistent with the ARPES study in Ref.~\cite{Teng2022} reporting a significant shift of the chemical potential at temperatures above $T_{\mathrm{C}}$. Aided by band-resolved optical conductivity calculations given in the Supplemental Material \cite{SM}, the variation of the low-energy spectral weight can be related to changes of the transition energy between multiple linear Fe 3$d$-bands along $K \rightarrow \Gamma$, $\Gamma \rightarrow A$, and $A \rightarrow L$ [see Fig.~\ref{Fig2}~(e)]. Note that, for all calculations, the energy axis needs to be rescaled by 1/2.1 to achieve a good match with the experiment. A comparable rescaling factor of 1/1.6 was observed in ARPES measurements \cite{Teng2023}. Moreover, a similar discrepancy between theory and experiment is reported for other magnetic kagome metals, such as the $R$Mn$_6$Sn$_6$ family and Co$_3$Sn$_2$S$_2$, signaling a significant influence of electronic correlations \cite{Wenzel2022, Xu2020}. This effect is also observable in the difference between the experimental and DFT plasma frequencies, which can be used to gauge the strength of the electronic correlations as done in Fig.~S8. Our data evidence much stronger correlations in FeGe compared to the nonmagnetic kagome metals, \AVS\ and ScV$_6$Sn$_6$, and even compared to the magnetic Co$_3$Sn$_2$S$_2$.

Upon cooling below $T_{\mathrm{C}}$, several modifications in the electronic structure are clearly detected. First, the Fano resonance shifts to lower energies and significantly broadens upon approaching $T_{\mathrm{C}}$, indicating strong electron-phonon coupling. This behavior is highly unusual for the temperature evolution of phonon modes, which are expected to sharpen and blueshift as the lattice hardens with decreasing temperature (see Supplemental Material for a more detailed discussion on the Fano resonance \cite{SM}). Second, there is a notable enhancement in the spectral weight around 250~\cm\ and a gradual formation of a sharp absorption feature at 100~\cm, as displayed in Fig.~\ref{Fig2}~(f). However, no spectral weight transfer from low to high energies, i.e., no optical signature of a CDW gap, is observed. The absence of a gap opening in the optical spectra is consistent with the results of the Raman studies \cite{Wu2023}.

To further investigate the effect of lattice distortion below $T_{\mathrm{C}}$ on the electronic structure, we carry out DFT calculations of the band structure and optical conductivity for the low-temperature phase. Below $T_{\mathrm{C}}$, a $2 \times 2 \times 2$ superstructure as depicted in Fig.~\ref{Fig1}~(a), involving a tri-hexagonal (trh) distortion of the Fe-kagome planes was observed by X-ray diffraction (XRD) measurements \cite{Teng2022}. Additionally, a partial dimerization of Ge1 atoms was proposed by XRD and inelastic X-ray scattering studies on annealed samples, as well as DFT simulations \cite{Wang2023, Shao2023, Chen2023, Miao2023, Shi2023}. In both cases, however, the distortion of the kagome planes is on the order of $10^{-4}$~\AA\ and, thus, extremely small. 

\begin{figure*}
        \centering
       \includegraphics[width=2\columnwidth]{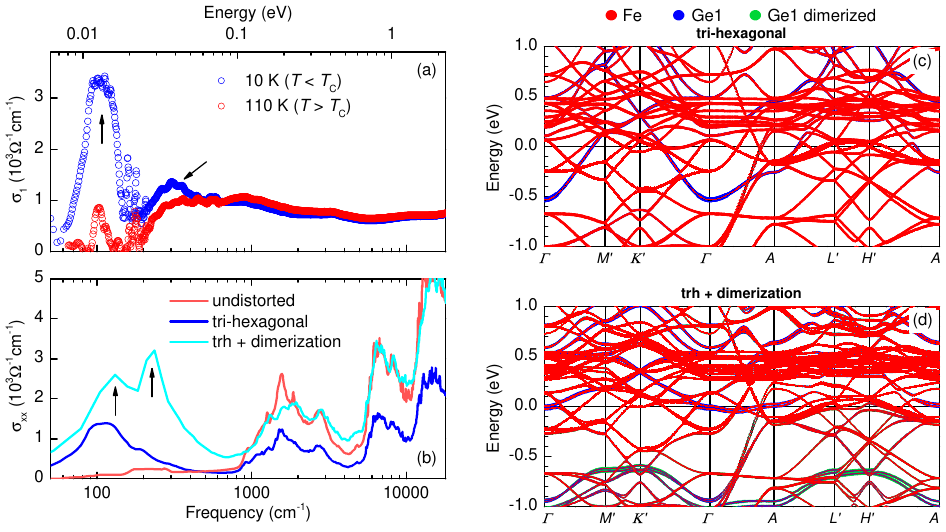}
   \caption{(a) Experimental interband transitions of FeGe at 110~K (normal state) and 10~K (low-temperature state). (b) Calculated optical conductivity given for the normal state, as well as for the simple tri-hexagonal distortion of the kagome planes and the Ge-dimerized superstructure. The arrows mark the two low-energy absorption bands observed below $T_{\mathrm{C}}$ in the experimental data, which are reproduced by the calculations only when taking into account both the tri-hexagonal distortion of the Fe kagome layers and the partial dimerization of the Ge1 atoms. Band structures of the low-temperature state for (c) the tri-hexagonal distortion of the kagome planes and (d) the Ge-dimerized superstructure. Colored dots show the contributions of different atoms. }	
    \label{Fig3}
\end{figure*}

The minuscule Fe displacement associated with the tri-hexagonal distortion induces only a tiny splitting between the parallel bands crossing the Fermi level (see Fig.~S7 for a more detailed view). Thus, plenty of spectral weight is found at low energies, resulting in the pronounced peak at 100~\cm\ as displayed in Fig.~\ref{Fig3}~(b). The splitting between some parallel bands can be increased by imposing a larger displacement of the Fe atoms. As discussed in the Supplemental Material \cite{SM}, this leads to more spectral weight at energies between 200 and 1000~\cm, with the low-energy peak at 100~\cm\ persisting. However, there is still no indication of a gap opening neither in the calculated optical conductivity nor in the density of states, as shown in Fig.~S6.  Hence, our study suggests a different nature of the low-temperature transition in FeGe. Moreover, experimental manifestations of this transition appear to be dependent on the growth method and require sensitive measurements as it was not detected in early studies \cite{Beckman1972, Bernhard1984}. Doping studies show a rapid suppression of the low-temperature anomaly in the magnetic susceptibility already at 1.5~\%~Sb-doping of the Ge-sites, militating in favor of the significant involvement of the Ge atoms in the low-temperature phase transition \cite{Huang2023}.

The Ge dimerization has a more drastic effect on the band structure compared to the deformation of the kagome planes. As seen in Fig.~\ref{Fig2}~(e) and Fig.~\ref{Fig3}~(c), precisely one in-plane Ge1 band crosses the Fermi level in the normal state. This band becomes split in the dimerized state with a gap of around 0.7~eV [panel~(d)]. While contributions from this splitting to the optical data are masked by several other interband optical transitions due to the multi-band nature of FeGe, a reorganization of the Fe bands near $E_{\mathrm{F}}$ leads to additional spectral weight at approximately 250~\cm, consistent with the experimental result (Figs. \ref{Fig3}(a) and \ref{Fig3}(b)). Along with the improved agreement between theory and experiment above 1000~\cm, these findings support the Ge dimerization scenario. Furthermore, the involvement of Ge1 atoms may also explain why the low-temperature transition has rather strong signatures in the optical spectra, i.e., Fano anomalies and different interband transitions, even though almost no changes happen in the Fe-kagome planes. 

The low-temperature anomaly in the magnetic susceptibility and the specific heat at $T_{\mathrm{C}}$ can be enhanced by a post-growth annealing process. However, the signature of the corresponding phase transition in the electric resistivity remains astonishingly weak \cite{Chen2023, Shi2023, Wu2023}. While additional Bragg peaks below $T_{\mathrm{C}}$ imply a structural modulation \cite{Teng2022}, this does not necessarily lead to a CDW. Kagome systems are prone to multiple structural instabilities, such as Wigner crystallization \cite{Chen2018, Wu2007}. Moreover, breathing distortion of the kagome lattice is a well-known effect in quantum magnetism, but also in the kagome metals where Fe$_3$Sn$_2$ adopts the breathing structure without any CDW \cite{Schaffer2017, Tanaka2020, Hirschberger2019, Regmi2022}.

Overall, there is a sharp contrast between FeGe and other kagome metals exhibiting a CDW at low temperatures, including the renowned \AVS\ compounds. Their kagome planes exhibit a very similar structural distortion at low temperatures, forming a 2~$\times$~2 in-plane superstructure. However, being on the order of $10^{-1}$~\AA, the deformation of the kagome layers is much more pronounced in the \AVS\ compounds \cite{Ortiz2021}. Interestingly, in their normal state, all these kagome metals feature multiple $d$-bands and exactly one $p$-band crossing the Fermi level. This $p$-band is related to the in-plane Ge1 atoms in FeGe but to out-of-plane Sb2 atoms in \AVS\ \cite{Tsirlin2022}. It essentially remains intact in the CDW phase of \AVS\ \cite {Ortiz2021} but splits in FeGe due to the partial dimerization of the in-plane Ge1 atoms. Lastly, our data do not reveal the presence of a gap in the low-temperature phase of FeGe, while clear signatures of a gap opening in ScV$_6$Sn$_6$ and the \AVS\ compounds were detected by infrared spectroscopy \cite{Wenzel2022a, Uykur2022, Uykur2021, Hu2023, Kim2023, Zhou2021, Zhou2023a}.

All these findings imply that the low-temperature phase of FeGe is substantially different from the CDW phase observed in other kagome metals. On the one hand, our study supports the partial dimerization of the in-plane Ge atoms, being decisive for the significant changes in the optical data below $T_{\mathrm{C}}$. On the other hand, the absence of a gap in both experiment and theory is clearly different from the conventional CDW scenario that has been discussed for FeGe previously. Given the complex interplay between magnetism, structural instabilities, strong electronic correlations, as well as the simple tunability of the low-temperature anomaly by doping and annealing \cite{Huang2023, Wu2023, Chen2023, Shi2023}, antiferromagnetic FeGe might open new routes to study so far unexplored regions in the rich phase diagram of kagome metals.

Authors acknowledge the fruitful discussions with S.~Fratini and the technical support by G.~Untereiner. M.W. is supported by Center for Integrated Quantum Science and Technology (IQST) Stuttgart/Ulm via a project funded by the Carl Zeiss Stiftung. The work has been supported by the Deutsche Forschungsgemeinschaft (DFG) via DR228/51-3, DR228/68-1, and UY63/2-1. 

\FloatBarrier
\nocite{Richardson1967, Homes1993, Tanner2015, wien2k, Blaha2020, pbe96, Kresse1996, Draxl2006, Fratini2014, Fratini2023, Kawamura2019, Shao2020}
\bibliography{FeGe}

\end{document}


\widetext
\begin{center}
\textbf{\large Supplemental Material for ''Intriguing Low-Temperature Phase in the Antiferromagnetic Kagome Metal FeGe''}

\vspace{5mm}
M. Wenzel, E. Uykur, A. A. Tsirlin, S. Pal, R. Mathew Roy, C. Yi, C. Shekhar, C. Felser, A. V. Pronin, M. Dressel
\end{center}
\setcounter{equation}{0}
\setcounter{figure}{0}
\setcounter{table}{0}
\setcounter{page}{1}
\makeatletter
\renewcommand{\theequation}{S\arabic{equation}}
\renewcommand{\thefigure}{S\arabic{figure}}
\renewcommand{\bibnumfmt}[1]{[S#1]}
\renewcommand{\citenumfont}[1]{S#1}

\section{Crystal growth}
B35-type FeGe single crystals are cultivated via the chemical vapor transport method, which employs iodine as a transport agent \cite{Richardson1967}. High-purity Fe and Ge powders are combined and ground in an agate mortar bowl with a molar ratio of  Fe:Ge~$=$~1:1. The mixture and iodine ($\sim$~10~mg/cm$^3$) are then loaded into a quartz tube and sealed under high vacuum. After that, the tube is placed in a two-zone furnace with the source powder at 893~K and the crystallization region at 833~K. After one month of maintenance, shiny crystals in the form of hexagonal prisms are produced at the cold end.

\FloatBarrier
\section{Experimental details}
\begin{figure}
        \centering
       \includegraphics[width=1\columnwidth]{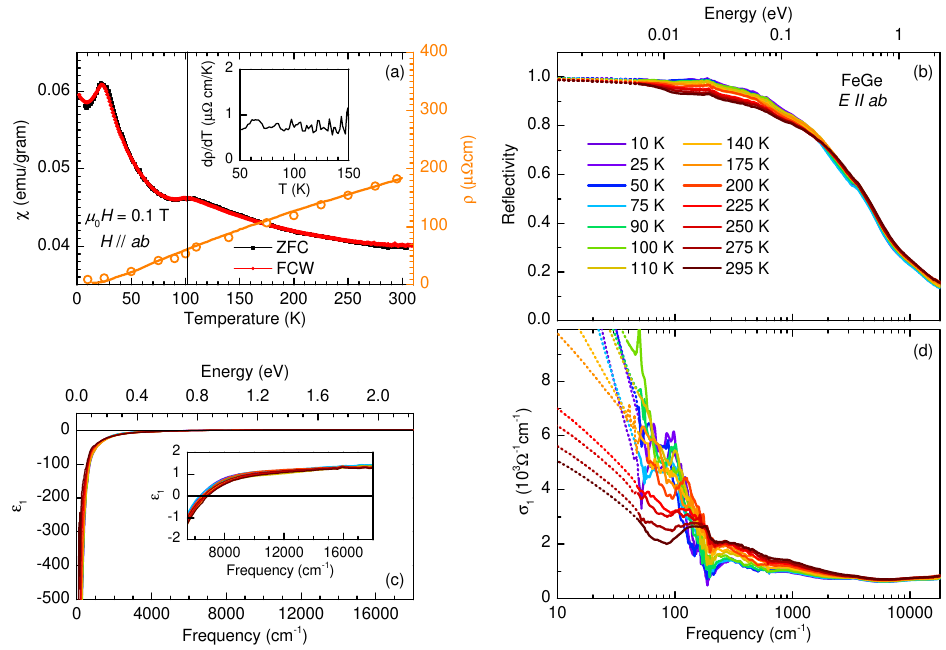}
   \caption{(a) Magnetic susceptibility and electric resistivity curves, measured in the $ab$-plane. The vertical black line marks the low-temperature anomaly observed in the magnetic susceptibility. Open circles are the dc resistivity values obtained from the Hagen-Rubens fits of the optical measurements. RRR~=~51 indicates the good quality of our sample. The inset displays the first derivative of the resistivity. (b) Temperature-dependent reflectivity over a broad frequency range measured in the $ab$-plane. Dotted lines are the Hagen-Rubens extrapolation to low energies. (c) Calculated dielectric permittivity. The inset highlights the zero crossing, marking the screened plasma frequency. (d) Temperature-dependent in-plane optical conductivity with the dotted lines indicating the Hagen-Rubens extrapolations to low energies.}
    \label{ROC}
\end{figure}

Temperature- and frequency-dependent reflectivity measurements in the $ab$-plane were performed on a freshly cleaved sample with the lateral dimensions of $\sim 0.5 \times 1~\mathrm{mm}^2$ and thickness of about 400~$\mu$m, as presented in Fig.~\ref{ROC}~(a). For high energies ($600\, \mathrm{cm^{-1}} \leq \omega \leq 18000$~\cm), a Vertex80v spectrometer coupled to a Hyperion IR microscope was used, while the low-energy range ($50\, \mathrm{cm^{-1}} \leq \omega < 600$~\cm) was measured with an IFS113v spectrometer and a custom-built cryostat. Freshly evaporated gold mirrors served as the reference throughout the measurements. The absolute value of the reflectivity is deduced with the gold-overcoating technique as described in Ref.~\cite{Homes1993} in the far-infrared range.

The optical conductivity is calculated from the measured in-plane reflectivity spectra employing standard Kramers-Kronig analysis. Below 50~\cm, the Hagen-Rubens relation is utilized for the data extrapolation, while x-ray scattering functions are used for the high-energy extrapolations \cite{Tanner2015}.

Prior to the optical measurements, the low-temperature anomaly at $T_{\mathrm{C}} = 102$~K was confirmed by dc magnetic susceptibility measurements in zero field cooling (ZFC) and field cooling (FC) regimes with $\mu_{\mathrm{0}} H = 0.1$~T. Additionally, four-point resistivity measurements were performed on the same sample piece as used for the optical experiments. The obtained curve matches well with the dc resistivity values obtained from the Hagen-Rubens fits as shown in Fig.~\ref{ROC}~(a). 

\section{DFT calculations}
Band structure and optical conductivity calculations were performed in the \texttt{WIEN2K} code \cite{wien2k, Blaha2020} using the Perdew-Burke-Ernzerhof flavor of the exchange-correlation potential \cite{pbe96}. Spin-orbit coupling was included in all calculations. Experimental magnetic configuration was used, with ferromagnetic kagome layers coupled antiferromagnetically. For the normal state, experimental lattice parameters and atomic positions from Ref.~\cite{Teng2022} were used, whereas the dimerized structure was taken from Ref.~\cite{Shao2023} where it has been obtained by a structural relaxation in the \texttt{VASP} code \cite{Kresse1996} using a cut-off energy of 520~eV. Experimental low-temperature structure with the tri-hexagonal distortion was taken from Ref.~\cite{Teng2022}. The dimerization was added to it using the Ge1 positions from Ref.~\cite{Shao2023}. Self-consistent calculations were converged on the $k$-mesh with 21~$\times$~21~$\times$~11 points for the normal state and 12~$\times$~12~$\times$~13 points for the low-temperature state. Optical conductivity was calculated using the \texttt{optic} module of \texttt{WIEN2K} \cite{Draxl2006} on the dense $k$-mesh with up to 58~$\times$~58~$\times$~29 points (normal state) and 36~$\times$~36~$\times$~38 points (low-temperature state). $\Gamma$-point phonons were calculated in \texttt{VASP} using the frozen-phonon method.

\section{Decomposition}
Different contributions to the total optical conductivity are modeled with the Drude-Lorentz approach. The dielectric function [$\tilde{\varepsilon}=\varepsilon_1 + i\varepsilon_2$] is expressed as
\begin{equation}
\label{Eps}
\tilde{\varepsilon}(\omega)= \varepsilon_\infty - \frac{\omega^2_{p,{\rm Drude}}}{\omega^2 + i\omega/\tau_{\rm\, Drude}} + \sum\limits_j\frac{\Omega_j^2}{\omega_{0,j}^2 - \omega^2-i\omega\gamma_j},
\end{equation}
with $\varepsilon_{\infty}$ being the high-energy contributions to the real part of the dielectric permittivity. As plotted in Fig.~\ref{ROC}~(c), these contributions are temperature-independent and approach the value of approximately 1.3. The other parameters $\omega_{p,{\rm Drude}}$ and $1/\tau_{\rm\,Drude}$ are the plasma frequency and the scattering rate of the itinerant carriers, respectively. Lorentzians described by the resonance frequency $\omega_{0,j}$, the strength of the oscillation $\Omega_j$, and the width $\gamma_j$ are used to model the interband absorptions.

The complex optical conductivity [$\tilde{\sigma}=\sigma_1 + i\sigma_2$] is calculated as
\begin{equation}
\tilde{\sigma}(\omega)= -i\omega \varepsilon_{\mathrm{0}}[\tilde{\varepsilon} (\omega) -\varepsilon_\infty],
\end{equation}
with $\varepsilon_{\mathrm{0}}$ being the vacuum permittivity.

Optical spectra of kagome metals are characterized by a broad absorption peak at low energies, which is intraband in nature, signaling the presence of charge carriers with hindered dynamics. In addition to the classical Drude-Lorentz approach, we use the displaced Drude peak formalism proposed in 2014 by Fratini \textit{et al.} \cite{Fratini2014} to model this so-called localization peak. Here, interactions of charge carriers with low-energy degrees of freedom, such as phonons and electric or magnetic fluctuations, can lead to a backscattering of the electrons, expressed by the backscattering time $\tau_ {\mathrm{b}}$. The resulting localization effect shifts the classical Drude peak at zero frequency to a finite value.
\begin{equation}
\tilde{\sigma}_{\rm localization}(\omega)=\frac{C}{\tau_{\mathrm{b}}-\tau}\frac{\tanh\{\frac{\hbar\omega}{2k_{\mathrm{B}}T}\}}{\hbar\omega}\,\cdot\,\mathrm{Re}\left\{\frac{1}{1-\mathrm{i}\omega\tau}-\frac{1}{1-\mathrm{i}\omega\tau_{\mathrm{b}}}\right\}.
\end{equation}
Here, $C$ is a constant, $\hbar$ is the reduced Planck constant, $k_{\mathrm{B}}$ the Boltzmann constant, and $\tau$ the elastic scattering time of the standard Drude model. At low temperatures, this peak becomes too sharp to be fitted with the displaced Drude peak. Hence, a regular Lorentzian is used to model this absorption.

The asymmetric phonon mode, at approximately 190~\cm, is reproduced most accurately by the Fano model
\begin{equation}
\sigma_1(\omega) = \frac{\sigma_0\omega\gamma[\gamma\omega (q^2-1)+2q(\omega^2-\omega_0^2)]}{4\pi[(\omega^2-\omega_0^2)^2+\gamma^2\omega^2]} \quad .
 \label{fano}
\end{equation}
This model can be considered a modified Lorentzian with an additional parameter, $q$. This dimensionless coupling parameter determines the coupling to the electronic background, leading to the asymmetry of the mode. 

The total optical conductivity takes the form
\begin{equation}
\tilde{\sigma}(\omega)= \underbrace{\tilde{\sigma}_{\mathrm{Drude}} + \tilde{\sigma}_{\mathrm{localization}}}_{\mathrm{intraband}} + \underbrace{\tilde{\sigma}_{\mathrm{Fano}}}_{\mathrm{phonon}} + \underbrace{\tilde{\sigma}_{\mathrm{Lorentzians}}}_{\mathrm{interband}}.
\end{equation}

\begin{figure}
        \centering
       \includegraphics[width=0.8\columnwidth]{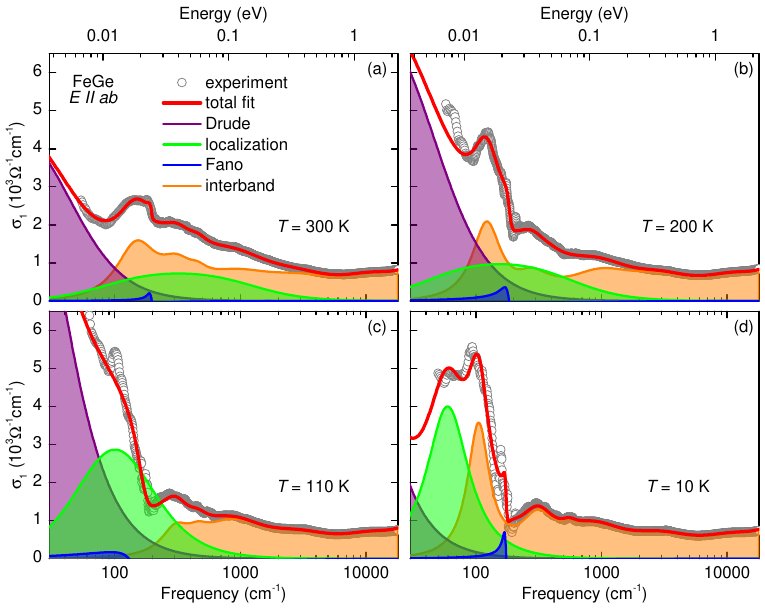}
   \caption{Decomposed optical conductivities at several temperatures above and below the low-temperature anomaly, consisting of a Drude peak (purple), a localization peak (green), a Fano resonance (blue), and several interband transitions (orange) modeled with the Drude-Lorentz approach.}	
    \label{decomp}
\end{figure}

To ensure the reliability of the fit parameters, the reflectivity, and the complex optical conductivity are fitted simultaneously. Fig.~\ref{decomp} displays the decomposed real part of the optical conductivity at selected temperatures. The temperature evolution of the localization peak and the Fano antiresonance are discussed in the upcoming sections in more depth.

\section{Localization peak}
The occurrence of a localization peak is a common feature among kagome metals. Upon cooling, this peak shows a pronounced linear redshift and may eventually merge with the conventional Drude peak at low temperatures, but also a saturation of the peak position at finite frequencies has been observed in some of the compounds. In the ferromagnetic Fe$_3$Sn$_2$, the onset of saturation coincides with a reorientation of the magnetic moments \cite{Biswas2020}. On the other hand, an interplay with a phonon mode was deemed to be the most likely scenario for the observed saturation in ferromagnetic GdMn$_6$Sn$_6$ \cite{Wenzel2022}. Moreover, a pinning of the localization peak was also reported in the non-magnetic KV$_3$Sb$_5$, where the CDW transition temperature coincides with the onset of saturation \cite{Uykur2022}.

\begin{figure}
        \centering
       \includegraphics[width=1\columnwidth]{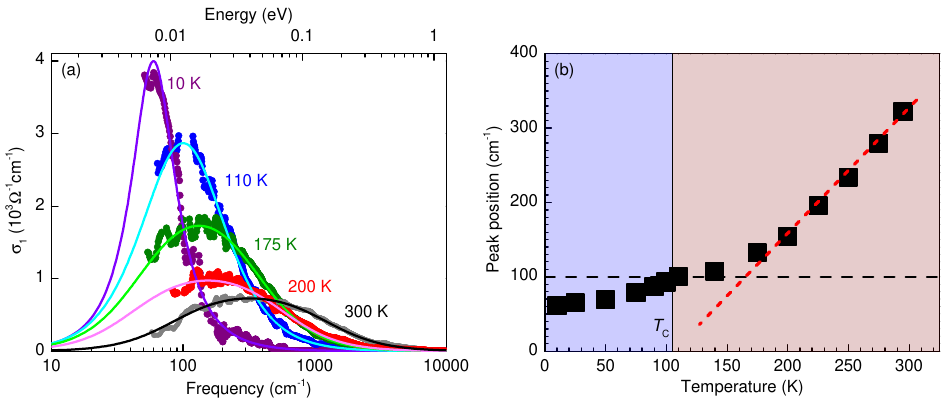}
   \caption{(a) Temperature dependence of the localization peak shown after subtracting all other contributions from the experimental optical conductivity. Solid lines are the Fratini model fits. (b) Peak position as a function of temperature. The red- and blue-shaded areas separate the high- and low-temperature phases. The red dashed line demonstrates the linear redshift at high temperatures, while the black dashed line marks the saturation frequency of the high-temperature phase. }	
    \label{loc}
\end{figure}

The temperature evolution of the localization peak in FeGe is given in Fig.~\ref{loc}. A linear redshift is observed at high temperatures, indicated by the red dashed line in panel (b). Below approximately 175~K, a deviation from the linear shift is detected, with the peak position saturating around 100~\cm. Given the absence of any magnetic or structural transition at such high temperatures, the most plausible explanation of such a behavior is the involvement of a phonon mode. 

Indeed, in the framework of the displaced Drude peak (DDP) formalism by Fratini \textit{et al.}, the localization peak would shift to lower energies upon cooling and pin to $\omega_0$ due to interactions between the localized carriers and a static bosonic mode at $\omega_0$ \cite{Fratini2014, Fratini2023}. Moreover, as temperature decreases, a sharpening and gain of intensity are predicted by theory and indeed observed experimentally in FeGe. Below $T_{\mathrm{C}}$, the localization peak becomes so sharp that it cannot be modeled with the displaced Drude peak anymore. Hence, we used a standard Lorentzian to fit the low-temperature data. A comparable sharpening of the localization peak was also reported in Fe$_3$Sn$_2$ at low temperatures \cite{Biswas2020}.

Below $T_{\mathrm{C}}$, the peak shifts to slightly lower energies. This shift may be related to a modification of the phonon energy due to the structural distortion below $T_{\mathrm{C}}$, assuming a phonon-mediated scenario. On the other hand, the sudden sharpening of the Fano resonance below $T_{\mathrm{C}}$ (seen Fig.~\ref{fano}~(e)) indicates a weakening of the electron-phonon coupling in the low-temperature phase. Moreover, assuming the coupling to phonons, the spectral weight of the localization peak can be used as a gauge of the electron-phonon coupling strength \cite{Wenzel2023}. As seen in Fig.~\ref{SW}~(a), this spectral weight is significantly reduced below $T_{\mathrm{C}}$, suggesting a reduction of electron-phonon coupling in the low-temperature phase.

\section{Fano resonance}
Fig.~\ref{fano}~(b) displays the Fano resonance, obtained after subtracting the intra- and interband contributions from the optical conductivity. The observed Fano mode around 190~\cm\ shows significant changes across $T_{\mathrm{C}}$. We note that the mode is an antiresonance, and consequently, the coupling parameter $q$ takes negative values. The strong anomalies become even more apparent when looking at the fit parameters given in panels (c)-(f). 

\begin{figure}[h]
        \centering
       \includegraphics[width=0.7\columnwidth]{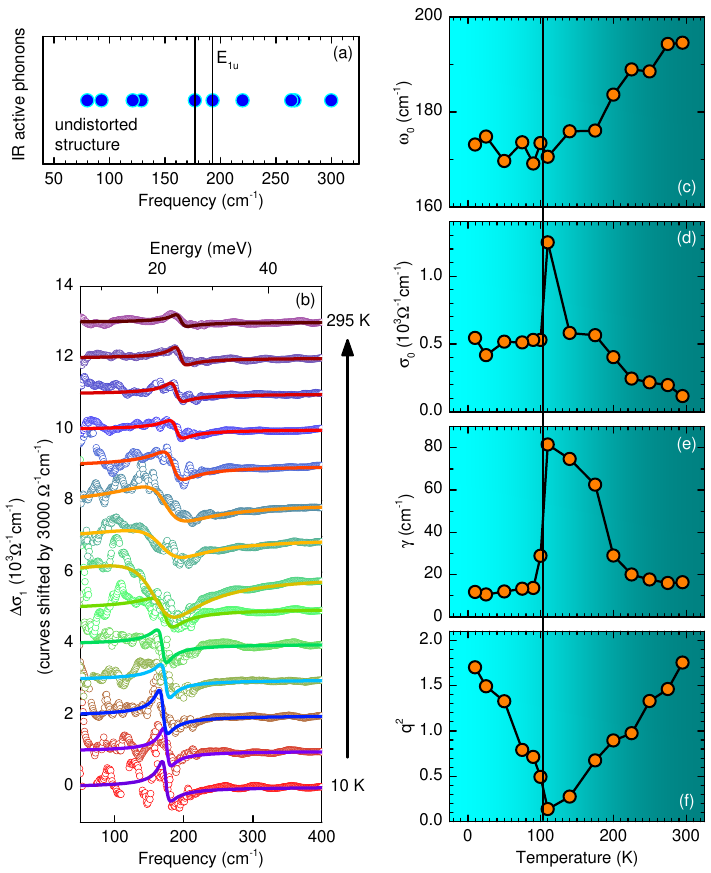}
   \caption{(a) Infrared-active phonon frequencies calculated for the undistorted structure. Vertical lines mark the two $E_{1u}$ modes with a comparable energy to the experimentally observed Fano resonance. (b) Low-energy optical conductivity, highlighting the Fano antiresonance. The spectra are presented as $\Delta \sigma_{\mathrm{1}}(\omega)$ with all other contributions subtracted and the baselines shifted by 3000~$\Omega^{-1}$cm$^{-1}$ for clarity. The solid lines are fits to the optical conductivity with the fit parameters given in (c)–(f). The vertical black line marks the anomaly at $T_{\mathrm{C}} = 102$~K. }	
    \label{fano}
\end{figure}

At high temperatures, a redshift of the resonance frequency $\omega_{\mathrm{0}}$ is accompanied by an increase of the intensity $\sigma_{\mathrm{0}}$. Most surprisingly, the Fano resonance reveals a notable broadening upon cooling, which is highly unexpected with decreasing thermal effects. The coupling constant, which also determines the asymmetry of the resonance, indicates a decreasing asymmetry upon cooling with the minimum at $T_{\mathrm{C}}$. Upon further cooling, the mode becomes more asymmetric and suddenly sharpens again. Together with the almost constant values of the resonance frequency below $T_{\mathrm{C}}$, the temperature evolution of $\omega_{\mathrm{0}}$ strongly resembles the behavior of the localization peak, further supporting the coupling scenario between localized carriers and phonons.

The occurrence of an asymmetric phonon mode in the optical spectra of kagome metals is not exclusive to FeGe. A Fano resonance was also reported in the optical spectra of ScV$_6$Sn$_6$, KV$_3$Sb$_5$, and RbV$_3$Sb$_5$, all three being kagome metals with a CDW phase at low temperatures. In ScV$_6$Sn$_6$ and FeGe, this mode sits at fairly low energies ($\sim$~190~\cm) and can be associated with one of the infrared active $E_{1u}$ modes according to our DFT calculations of FeGe given in Fig.~\ref{fano}~(a). In contrast, the interpretation of the Fano resonance in the \AVS\ compounds is not so straightforward, as it sits at much higher energies, where no more IR-active phonons are expected \cite{Uykur2022, Wenzel2022a}.

\section{Band-resolved optical conductivity}
\begin{figure}[h]
        \centering
       \includegraphics[width=0.85\columnwidth]{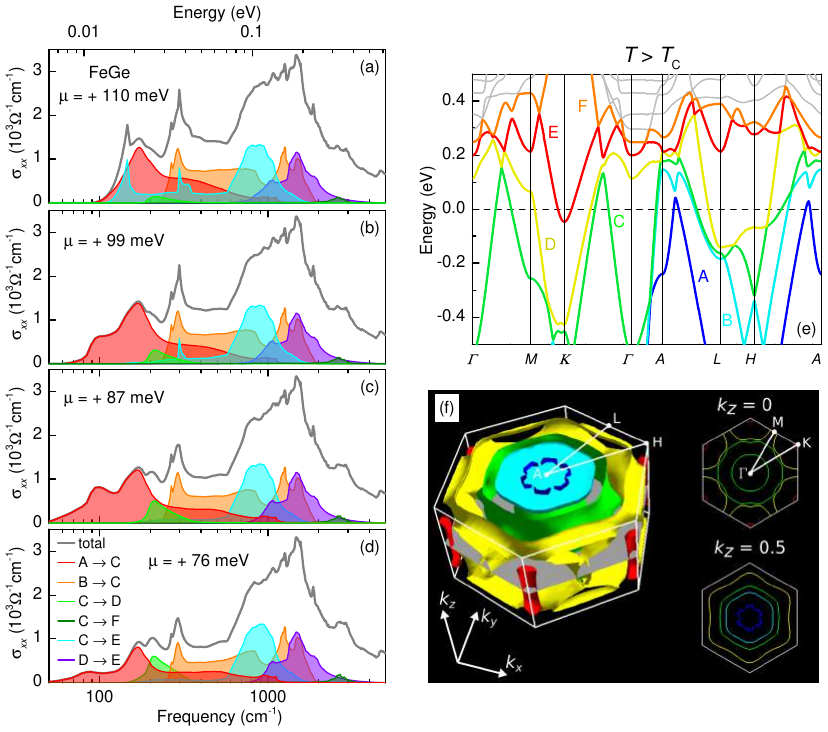}
   \caption{(a)-(d) Band-resolved optical conductivity given for different electron dopings expressed by the chemical potential $\mu$. Colors represent contributions from interband optical transitions between different bands according to the labeling in panel (e), showing the band structure for the normal state. (f) Fermi surface constructed from the calculated band structure and cuts along $k_z = 0$ and $k_z = 0.5$. \texttt{FermiSurfer} program was used for the visualization \cite{Kawamura2019}.}	
    \label{bandselect}
\end{figure}

As discussed in the main text, the highly temperature-sensitive interband transitions with $\omega \leq 200$~\cm\ above $T_{\mathrm{C}}$ are one of the key features in the optical spectra of FeGe. We have performed band-resolved optical conductivity calculations to relate these low-energy absorptions with specific transitions in the band structure. Due to the large amount of bands crossing $E_{\mathrm{F}}$, the Fermi surface, given in Fig.~\ref{bandselect}~(f), is highly sensitive to the position of the chemical potential. Electron doping results in an upward shift of the chemical potential, allowing transitions at very low energies. As seen in Fig.~\ref{bandselect}~(a-d), these low-energy absorptions are mainly related to transitions between bands C and D, following the labeling in panel (e). Along $K \rightarrow \Gamma$, these bands are almost parallel and become closer when shifting the chemical potential upwards.

\section{Additional computational results}
\begin{figure}[h]
        \centering
       \includegraphics[width=0.98\columnwidth]{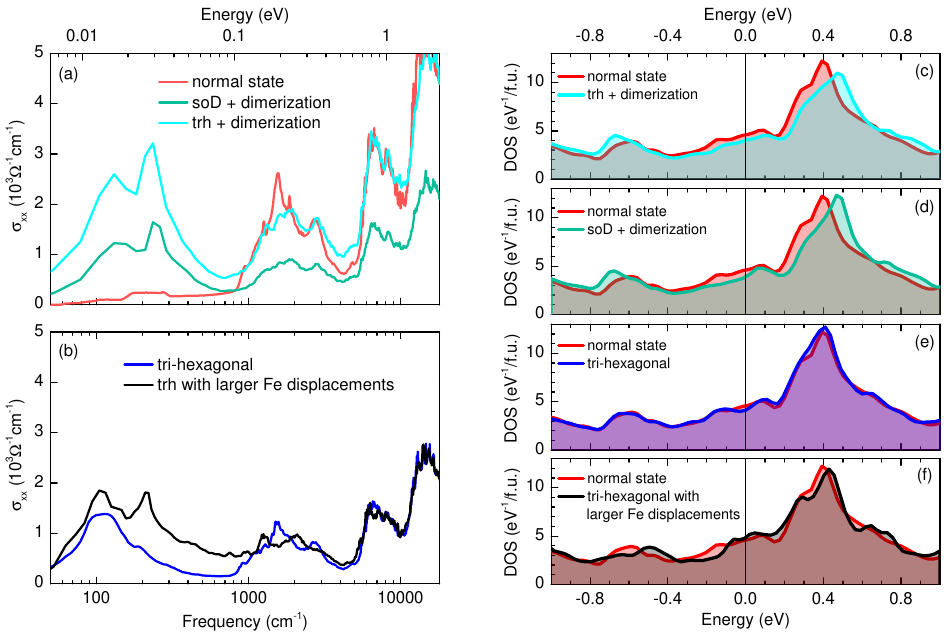}
   \caption{(a) Comparison of calculated optical conductivities for the dimerized superstructures with a tri-hexagonal and star-of-David-type distortion of the Fe-kagome planes. (b) Calculated optical conductivities for the tri-hexagonal in-plane distortion with different Fe displacements as explained in the text. (c-f) Corresponding density of states overlapped with the density of states for the normal state without shifting the chemical potential}.	
    \label{DOS}
\end{figure}

The absence of a gap in the experimental and the theoretical optical conductivity is also paralleled by the density of states (DOS) given in Fig.~\ref{DOS}~(c) and (e) for the dimerized and simple tri-hexagonal distortion, respectively. For both superstructures, only a small reduction in the number of states near $E_{\mathrm{F}}$ is found. In addition to the tri-hexagonal distortion of the kagome layers, a star-of-David-type (soD) distortion as observed in CsV$_3$Sb$_5$ could be envisaged \cite{Shao2023, Ortiz2021}. With the partially dimerized Ge atoms, the optical conductivity of the soD kagome distortion resembles the conductivity in the case of the tri-hexagonal distortion with overall smaller absolute values as shown in Fig.~\ref{DOS}~(a). Given the better agreement between the optical conductivities in the normal state and in the case of the tri-hexagonal distortion at high energies, we deem the soD-type distortion of the kagome planes less likely in FeGe.

\begin{figure}[h]
        \centering
       \includegraphics[width=1\columnwidth]{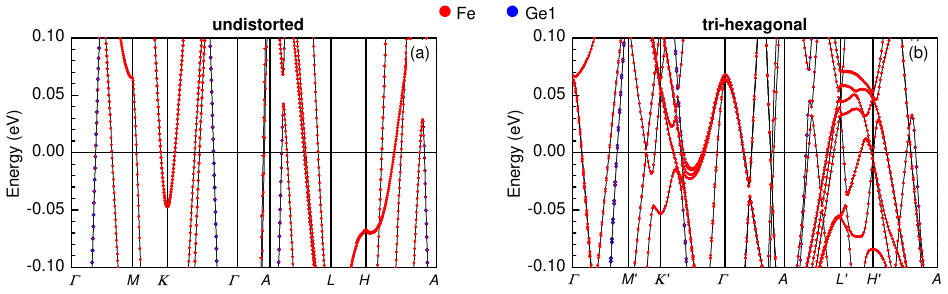}
   \caption{Band dispersions nearby the Fermi level for the normal state (a) and the tri-hexagonal distortion (b). Colored dots show contributions of the Fe atoms (red) and Ge1 atoms (blue). }	
    \label{splitting}
\end{figure}

In comparison with other kagome metals exhibiting charge order, the in-plane distortion of the Fe-kagome layers detected by XRD measurements is extremely small \cite{Teng2022}. Hence, the main effect of this distortion is a tiny splitting between the parallel bands crossing the Fermi level, as highlighted in Fig.~\ref{splitting}. We further explore the influence of the kagome distortion by imposing a larger Fe displacement similar to the \AVS\ compounds \cite{Uykur2021}. This indeed increases the splitting between a few parallel bands crossing the Fermi level. However, only some additional spectral weight at frequencies above 100~\cm\ is produced, as shown in Fig.~\ref{DOS}~(b). Consequently, there is still no observable spectral weight transfer from low to high energies that would indicate the opening of a gap below $T_{\mathrm{C}}$. In fact, with two distinct low-energy absorptions at 100~\cm\ and 250~\cm, the calculated optical conductivity resembles the optical conductivities of the dimerized superstructures given in Fig.~\ref{DOS}~(a). The DOS in panel (f) verifies the absence of a gap as no reduction of the states near $E_{\mathrm{F}}$ is observed.

\section{Plasma frequency and electronic correlations}
The experimental plasma frequencies are obtained from the spectral weight (SW) analysis of the Drude and localization peaks according to
\begin{equation}
\omega_{\mathrm{p}} = \sqrt{\mathrm{SW_{Drude} + SW_{loc}}}.
\end{equation}
The spectral weight is calculated from the optical data by integrating the real part of the optical conductivity of the fitted Drude and localization peaks
\begin{equation}
\mathrm{SW} = \frac{1}{\pi^2\varepsilon_{\mathrm{0}}c} \int_0 ^{\omega_{\mathrm{c}}} \sigma_{\mathrm{1}}(\omega)\mathrm{d}\omega,
\end{equation}
with $c$ being the speed of light. The cut-off frequency is chosen as $\omega_{\mathrm{c}} = 50000$~\cm, considering the high-energy tail of the localization peak. 

\begin{figure}
        \centering
       \includegraphics[width=1\columnwidth]{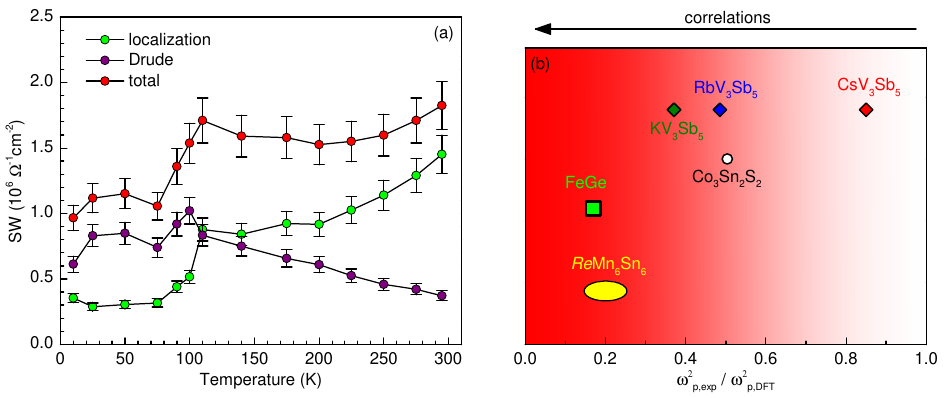}
   \caption{(a) Temperature evolution of the intraband spectral weight. (b) Strength of electronic correlations for different kagome metals taken from Refs.~\cite{Wenzel2022, Wenzel2022a, Shao2020}.}	
    \label{SW}
\end{figure}

The temperature evolution of the intraband spectral weight is given in Fig.~\ref{SW}~(a). Above $T_{\mathrm{C}}$, there is a spectral weight transfer between the Drude and the localization peaks, while the total SW can be considered constant within the error bars. Below 110~K, the spectral weight of both intraband contributions is reduced, leading to a drop in the total SW and, consequently, in the plasma frequency. The plasma frequency can be written as $\omega_{\mathrm{p}} \propto n/m^*$, with $n$ being the carrier density and $m^*$ the effective mass. Consequently, the decrease of $\omega_{\mathrm{p}}$ below $T_{\mathrm{C}}$ may be attributed to a reduction of the carrier density, signaling a gap opening. However, our calculations reveal that the carrier density at the Fermi level is not affected significantly by the structural transition at $T_{\mathrm{C}}$, ruling out the gap-opening scenario. Given the splitting between parallel bands crossing the Fermi level, as well as the splitting of the Ge1-band in the dimerized case, the effective mass is likely to be affected by the reconstruction of bands, leading to the reduction of the plasma frequency.


The large rescaling factor of the energy axis needed to match the DFT calculations with the experimental results indicates the presence of considerable electronic correlations. The strength of electronic correlations can be estimated by the ratio $\omega^2_{\mathrm{p, experiment}} / \omega^2_{\mathrm{p, DFT}}$ \cite{Shao2020}. Using the experimental plasma frequency at 110~K and the DFT calculations with the chemical potential shifted by +~76~meV, a value of approximately 0.17 is obtained. Note that this value is only slightly affected by the exact position of the chemical potential. Such a small value signals strong electronic correlations in the same range as observed for the $R$Mn$_6$Sn$_6$~($R =$~Gd, Tb) compounds \cite{Wenzel2022}. In contrast, relatively moderate to almost absent electronic correlations are found in the \AVS\ compounds \cite{Wenzel2022a}. 

\FloatBarrier
\bibliography{FeGe_SM}